\documentclass[authoryear,12pt]{elsarticle}
\usepackage{pdfpages}

\usepackage{import, graphicx, xcolor, calc}
\usepackage{subfigure}
\usepackage{hyperref}
\usepackage{float}

\usepackage{tikz}
\usetikzlibrary{shapes.geometric, arrows.meta, positioning, fit, backgrounds}

\tikzset{
    block/.style  = {rectangle, rounded corners, draw, minimum width=3cm, minimum height=0.8cm, 
                     align=center, font=\small},
    input/.style  = {rectangle, rounded corners, draw, minimum width=2.8cm, minimum height=1.4cm,
                     align=center, font=\small},
    output/.style = {rectangle, rounded corners, draw, minimum width=2cm, minimum height=0.8cm,
                     align=center, font=\small},
    arrow/.style  = {->, >=latex, thick},
    colbox/.style  = {rectangle, draw, dashed, inner sep=0.4cm, rounded corners}
}

\usepackage[fleqn]{amsmath}

\journal{Journal}

\floatstyle{plaintop}
\restylefloat{table}

\usepackage[ left=2.5cm, right=2.5cm, top=2.5cm, bottom=2.5cm, includehead]{geometry}

\newcommand{\dcl}[1]{ \multicolumn{2}{c}{ #1 } }

\begin{document}

\begin{frontmatter}
    \title{Orientation-Resolved Scaling Relations for Mass Transport and Pressure Drop in Spacer-Filled Reverse Osmosis Channels}

    \author{F. J. Aschmoneit$^1$\corref{mycorrespondingauthor}}
   \cortext[mycorrespondingauthor]{Corresponding author}
    \ead{fynnja@math.aau.dk}
   \author{C. H\'elix-Nielsen$^2$}
    \address{(1) Department of Mathematical Sciences, Aalborg University, Copenhagen, Denmark}
    \address{(2) Department of Environmental and Resource Engineering, Technical University Denmark, Kgs. Lyngby, Denmark}
    

\begin{abstract}
    Feed spacers in spiral-wound reverse osmosis modules enhance mass transport but increase the flow resistance.
    Both effects depend on the spacer orientation, i.e. the flow attack angle $\alpha$, which has so far been characterized only at selected angles.
    This work derives continuous scaling relations for the Sherwood number ($Sh$) and the friction factor ($f$) as functions of $\alpha$ and the Reynolds number ($Re$) from high-resolution CFD simulations of two commercial overlapped spacers.
    The coefficients are obtained by an analytic least-square fit.
    Mass transport follows $Sh = 6.65(1 + 0.244 \sin(2\alpha)) (d_h/L)^{1/2} Re^{1/2}$ and the flow resistance follows $f = 11.63(1 + 0.19 \sin(2\alpha)) Re^{-1/2}$ for $Re > 80$.
    The relations hold for overlapped spacers with circular filaments at $Sc \approx 670$ and $Re$ up to about 210.
    The effects of $Re$ and $\alpha$ are separable with one set of coefficients for both spacer sizes.
    Rotating the spacer from $0^\circ$ to $45^\circ$ increases $Sh$ by 24\% and $f$ by 19\% at constant cross-flow velocity.
    Combined with an algebraic water flux model, the relations give closed-form estimates of water flux and pressure drop.
    For typical seawater and brackish water conditions, orientation increases the water flux by only 1\% to 2\% because concentration polarization is weak, but the gain grows with membrane permeability.
\end{abstract}

\begin{keyword} scaling relations, mass transport, friction factor, flow orientation, spacer-channels, transport phenomena, computational fluid dynamics, reverse osmosis, algebraic water flux, efficiency optimization
\end{keyword}
\end{frontmatter}

\section{Introduction}

    The filtration efficiency of reverse osmosis (RO) systems with spiral-wound modules is determined by the filtration rate and the hydraulic pressure loss through the module.
    The filtration rate is limited by the concentration build-up at the membrane, the concentration polarization (CP), which is counteracted by lateral mass transport, whereas the pressure loss reflects the flow resistance.
    Both are primarily governed by the channel flow velocity and are conventionally quantified through the Sherwood number ($Sh$) and the friction factor ($f$), respectively, for which CFD of small spacer cells is widely used to derive correlations \citep{toh_review_2020, sreedhar_evolution_2023}.
    Their dependence on the Reynolds number ($Re$) follows power laws, $Sh \sim Re^n$ and $f \sim Re^{-m}$, with exponents that depend on the flow regime.
    For $Sh$, the reported exponent ranges between about $1/4$ and $2/3$ \citep{cespedes_improved_2025, gurreri_flow_2016, shi_performance_2015, koutsou_numerical_2009, fimbres-weihs_numerical_2007}.
    $f$ follows the laminar Darcy scaling $m=1$ at low Reynolds numbers \citep{mokhtar_cfd_2021, da_costa_spacer_1994}.
    In transitional flows, reported values are $m=1/2$ \citep{mokhtar_cfd_2021} or $m\approx1/3$ \citep{shi_performance_2015, farkova_pressure_1991}.\\

    Orientation affects both mass transport and pressure drop simultaneously.
    Two spacer angles must be distinguished: the filament angle between the two filament layers and the flow attack angle at which the feed approaches the spacer.
    In this work, the spacer orientation denotes the flow attack angle $\alpha$, i.e. the angle of filament alignment relative to the incoming flow direction, as shown in Fig. \ref{fig:SketchOfChannel}.
    The effect of the flow attack angle has been quantified at selected angles, mostly for woven spacers or in membrane distillation and reverse electrodialysis channels \citep{gurreri_flow_2016, ponzio_experimental_2017, albeirutty_experimental_2018, mokhtar_cfd_2021}.
    For commercial spacers, velocity fields and pressure losses have been measured at two orientations \citep{haidari_determining_2019}.
    These studies mostly report the best mixing at $45^\circ$, whereas the effect on the pressure drop differs between spacer types \citep{haidari_determining_2019, mokhtar_cfd_2021}.
    The orientation dependence is thus reported as trends at selected angles rather than as a quantitative relation across operating conditions and spacer sizes.
    For the filament angle, continuous correlations of the energy loss factor with $Re$ exist \citep{guan_hydrodynamic_2023, hu_enhanced_2025}.
    They are, however, restricted to specific spacer structures \citep{hu_enhanced_2025} and do not cover mass transfer.
    Models identified from CFD simulations predict the pressure drop over a wide range of filament thicknesses and orientations and reveal strongly nonlinear sensitivities to these parameters \citep{johannink_predictive_2015}.
    The spacer geometry also governs the pressure drop increase caused by biofouling \citep{bucs_effect_2014}.
    Mass transfer is, however, not covered by these models.
    Module-scale models account for the spacer through correlations in which angles enter only via porosity and hydraulic diameter \citep{al-obaidi_optimizing_2024}.
    To the authors' knowledge, a continuous relation for the dependence of both $Sh$ and $f$ on the flow attack angle has not been reported, which is the gap this work addresses.\\

    How much a gain in mass transport improves the water flux depends on the extent of concentration polarization, which increases with membrane permeability \citep{sreedhar_evolution_2023, liang_role_2025} \citep{luo_supercomputing_2023}.
    Coupling spacer-scale correlations with module-scale models has been recommended to assess spacer performance in practice \citep{liang_role_2025}.
    Recent multiscale frameworks couple data-driven models trained on spacer-scale CFD simulations with system-level models to optimize spacer geometry, membrane permeability and process design simultaneously \citep{luo_supercomputing_2023, chen_rapid_2024, jeong_cfd_2026}.
    In combination with the algebraic water flux framework \citep{tiraferri_standardizing_2023}, the scaling relations derived here allow a closed-form assessment of the trade-off between water flux and pressure drop.
    Such a relation makes the orientation a continuous design variable in process optimization, where the concave flux gain and the linearly increasing pressure drop may favour intermediate angles, in particular for highly permeable membranes with stronger concentration polarization.
    Explicit scaling relations can be implemented directly in system-level design tools, as for example in an updated version of OMSD \citep{aschmoneit_omsd_2020}.\\

    This article is organized as follows:
    Section \ref{sec:methods} presents the spacer geometry, the governing equations and the CFD setup, including the periodic formulation of the solute transport.
    Section \ref{sec:results} presents the simulation data, derives the orientation-resolved scaling relations and applies them to the water flux and the pressure drop in SWRO and BWRO.
    Section \ref{sec:discussion} discusses the results in the light of previous studies and states the limitations of the present work, followed by the conclusions.
    The analytic least-square fitting is detailed in \ref{app:leastSquare}.\\

    The scientific contributions of this article can be summarized as follows:

    \begin{itemize}
        \item An orientation-resolved scaling relation for $Sh$ and $f$ that collapses two commercial overlapped (non-woven) spacers onto a single relation in which the effects of $Re$ and $\alpha$ are separable.
        \item A closed-form screening scheme showing that the flux benefit of spacer orientation is governed by the ratio $K$ of mass transport to permeation and grows with membrane permeability.
    \end{itemize}

\section{Problem Formulation and Methods}\label{sec:methods}

\subsection{Spacer Geometry, Hydraulic Diameter and Dimensionless Groups}

\begin{figure}[H]
        \centering
        \includegraphics[width=0.7\linewidth]{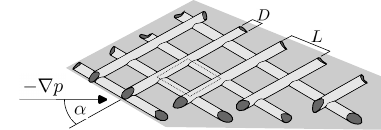}
        \caption{Sketch of regularly confined channel of height $H$ (in normal-direction of the gray plane). The cross-flow is driven by the pressure gradient $\nabla p$ at an incident angle $\alpha$. The dashed rectangle indicates the computational domain. $D$ and $L$ represent the rod diameter and the mesh spacing, respectively.}
        \label{fig:SketchOfChannel}
    \end{figure}

    Consider a spacer-channel of height $H$, which is constricted by a regular, perpendicular mesh of cylindrical rods with spacing $L$ and diameter $D= 3/5\,H$, creating an overlap at the rod crossings.
    The cross-flow is driven by the pressure gradient $\nabla p$ at an incident angle $\alpha$, see Fig \ref{fig:SketchOfChannel}.

    In non-circular, uniform channels the hydraulic diameter $d_h$ corresponds to the hypothetical diameter of a circular pipe with equivalent pressure drop.
    Hence, its value depends on the channel's cross-sectional geometry and is equated as four times the cross-sectional area, divided by the wet cross section perimeter.
    Similarly, in non-uniform channels, the hydraulic diameter is commonly defined as:
    
    \begin{equation*}
        d_h = \frac{4 V_\text{wet}}{A_\text{wet}},
    \end{equation*}

    where $V_\text{wet}$ is the water-filled volume, and $A_\text{wet}$ is its wetted boundary area.
    $V_\text{wet}$ is defined as the difference of the empty channel volume and the volume occupied by the rods.
    $A_\text{wet}$ is defined as the quadratic top and bottom boundaries, plus the area of the rods.
    In the case of intersecting rods, i.e. if $2D > H$, the intersection volume must be accounted for, which reduces both $A_\text{wet}$ and $V_\text{wet}$.
    The evaluation of the intersection volume and its surface area is a non-trivial mathematical exercise.
    In this article, approximate functions for the volume and surface area are used, see \cite{aschmoneit_volume_2025} for their derivation.
    With $\delta = 2-H/D$ denoting the relative intersection depth, $V_\text{wet}$ and $A_\text{wet}$ are defined as:
    
    \begin{equation*}
        V_\text{wet} = 
        \underbrace{\vphantom{\frac{1}{r}}HL^2}_\text{empty channel} 
        - \underbrace{\frac{\pi}{2} D^2 L}_\text{two rods} 
        - \underbrace{ \frac{D^3}{3} (1-cos(\delta \pi))}_\text{intersection}
    \end{equation*}

    \begin{equation*}
        A_\text{wet} = 
        \underbrace{\vphantom{\frac{1}{r}}2L^2}_\text{top, bottom} 
        + \underbrace{\vphantom{\frac{1}{r}}2\pi D L}_\text{two rods}
        - \underbrace{ 4 D^2 \sin \left( \frac{\delta \pi}{2} \right) }_\text{intersection}        
    \end{equation*}
    
    Table \ref{tab:spacerConfigurations} summarizes the geometric properties of the two spacers investigated in this study. 
    They relate to mass transport enhancing feed spacers in commercially available reverse osmosis filtration systems.

    \begin{table}[H]
    \begin{center}
    \begin{tabular}{ ccccccc } 
     \hline
          \textbf{spacer}      & $\mathbf{L} \, [\mu m]$     & $\mathbf{H} \, [\mu m]$ & $\mathbf{D} \, [\mu m]$ & $\mathbf{\delta}$   &  $\mathbf{d_h} \, [\mu m]$ & $\mathbf{d_h / L}$ \\ 
          \hline
     28 mil  & 2822 & 711 & 427 & 1/3 & 838 & 0.297  \\ 
     34 mil  & 3629 & 867 & 520 & 1/3 & 1046 & 0.288 \\ 
     \hline
    \end{tabular}
    \caption{Spacer geometry specs}\label{tab:spacerConfigurations}
    \end{center}
    \end{table}

    The simulation results are represented through the dimensionless Sherwood number $Sh$, the Darcy friction factor $f$ and the Reynolds number $Re$, with
        \begin{equation*}
            Sh = \frac{d_h}{\Delta c} \frac{\partial c}{\partial n},\quad f  = \frac{8 \, \tau}{\rho U^2}, \quad Re = \frac{U d_h}{\nu}
        \end{equation*}

\subsection{Governing Equations and Periodic Solute Transport}

    The flow through the spacer-filled channel is driven by a constant pressure gradient in the $-\hat{e}_x$ direction, which imposes a bulk flow in the $\hat{e}_x$ direction.
    Hence, the pressure gradient has a constant component $-\beta$ in the $\hat{e}_x$ direction:
    \begin{equation}
    \label{eq:pressureGradient}
    \nabla P = \nabla p - \beta \, \hat{e}_x,
    \end{equation}
    Since the flow is laminar, $Re < 250$, it assumes a fully developed profile that is shaped by the periodically repeating spacer filaments.
    Furthermore, the flow is assumed steady, incompressible and Newtonian. 
    Direct numerical simulations of a commercial non-woven spacer show steady flow at a cross-flow velocity of $0.185\,m\,s^{-1}$ \citep{qamar_novel_2021}.
    Under these assumptions, the Navier-Stokes equations for the velocity $\mathbf{u}$ are:

    \begin{equation*}
    (\mathbf{u} \cdot \nabla)\, \mathbf{u} = - \frac{1}{\rho} \nabla p + \nu \nabla^2 \mathbf{u} + \frac{\beta}{\rho}\, \hat{e}_x
    \end{equation*}
    Assuming further that the solute concentration has a negligible effect on the water density, mass conservation is given by the continuity equation:
    \begin{equation*}
    \nabla \cdot \mathbf{u} = 0
    \end{equation*}
    The solute concentration is treated as a passive scalar, i.e. it does not affect the flow field.
    Its transport is then governed by the steady velocity field and the diffusion coefficient $D$:
    \begin{equation*}
    \mathbf{u} \cdot \nabla s - D \nabla^2 s = 0
    \end{equation*}

    While the velocity field is periodic across the spacer cell, the solute concentration increases along the feed channel and is therefore not periodic.
    This is circumvented by a normalized, periodic solute concentration, analogous to the normalized, periodic temperature that \cite{patankar_fully_1977} define for heat transport in periodically changing ducts:

    \begin{equation*}
    s = \frac{c(x,y) - c_w}{c_b(x) - c_w},
    \end{equation*}
    
    where $c(x,y)$ is the local concentration, $c_w$ is the constant wall concentration and  $c_b$ is a reference concentration.
    Treating $s$ as a periodic variable, and since $\partial_n s \sim \partial_n c$, there is no need to convert the relative concentration into an absolute concentration. 
    The solute mixing is directly measured from the periodic concentration.

\subsection{Numerical Setup and Verification}

\begin{figure}[H]
        \centering
        \includegraphics[width=0.95\linewidth]{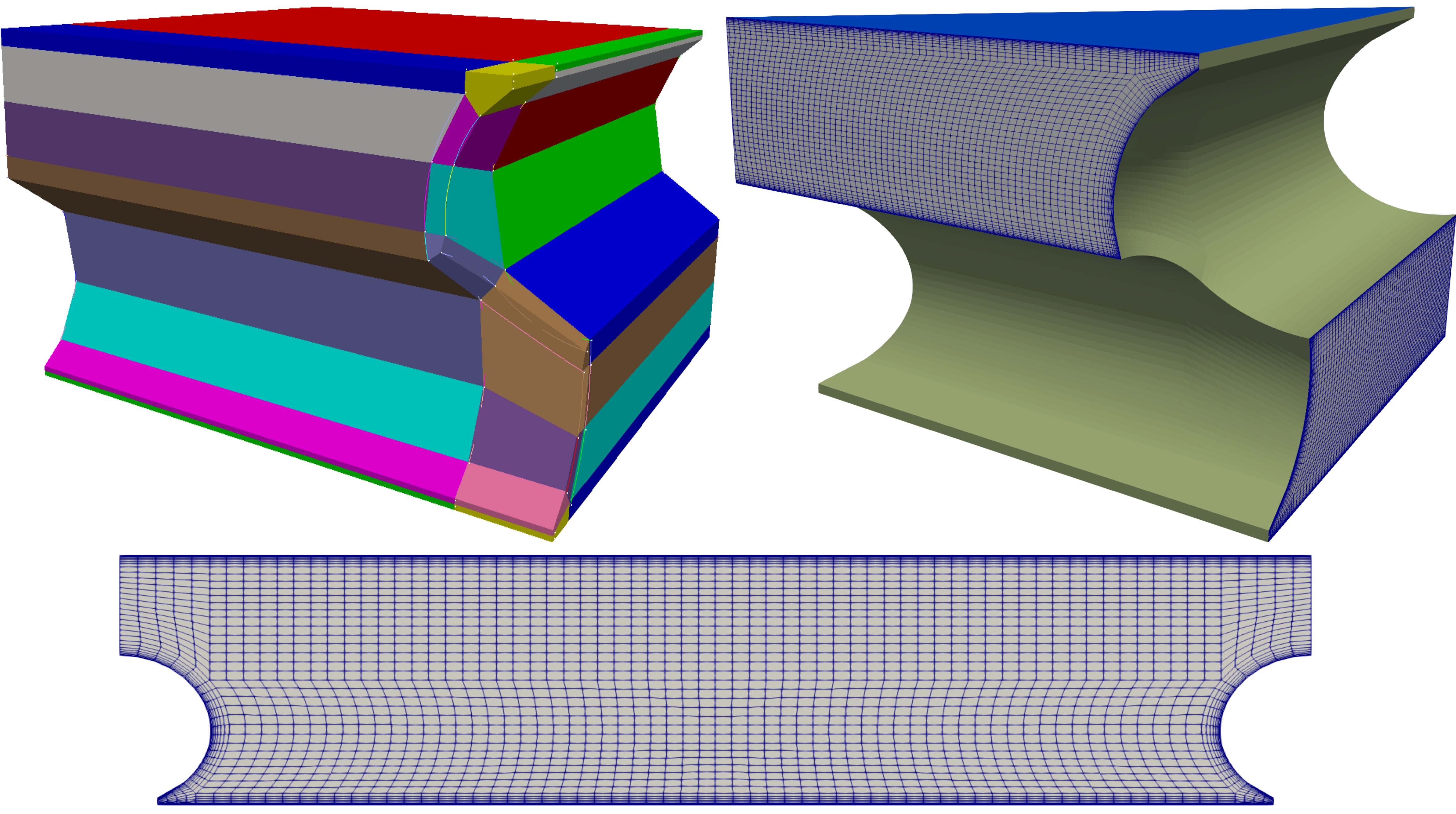}
        \caption{Computational mesh at intersection of traversing rods, with $\delta = 1/3$. Top: Block arrangement for OpenFOAM's blockMesh. Middle: Resulting mesh around intersection. Bottom: Cross section of channel indicating boundary layer resolution.}
        \label{fig:blocksAndMesh}
    \end{figure}

    The confined channel is decomposed into periodic cells of rod segments, see dashed rectangle in Fig. \ref{fig:SketchOfChannel}.
    The computational mesh is created using OpenFOAM's 'blockMesh' tool\footnote{\url{https://gitlab.gbar.dtu.dk/fyna/ambrosia-diamondBaseCase}}.
    Exploiting the two symmetry planes in the geometry reduces the mesh development to one quarter of the domain, which is then mirrored across the symmetry planes. 
    Figure \ref{fig:blocksAndMesh} (top) illustrates the block design used for mesh generation. 
    The periodic cell geometry is discretized using hexahedral cells, with boundary layers of increased resolution near the spacer and membrane surfaces, as shown in Fig. \ref{fig:blocksAndMesh}.

    Following the procedure of \citet{celik_procedure_2008}, a Grid Convergence Index (GCI) study was performed on the 28 mil spacer geometry to quantify spatial discretization error convergence, with the wall-normal scalar concentration gradient (Eq. \ref{eq:wallNormalScalarGrad}) as target quantity, which is most sensitive to near-wall mesh resolution.
    The results suggested that the ideal mesh resolution for the 28 mil geometry could also be applied to the 34 mil geometry. 
    The GCI simulations were conducted with a prescribed pressure gradient to achieve a bulk flow velocity of approximately $20 \, \text{cm/s}$ and a mesh orientation of $45^\circ$, representing the most demanding case for near-wall resolution: 
    the highest tested $Re$ yields the thinnest concentration boundary layer, while $45^\circ$ produces the most complex local mixing pattern.
    A bulk discretization of $140 cells/\text{mm}$ and 14 boundary layers with a growth rate of 1.1 resulted in a discretization error of less than 1\% in this target quantity. 
    For the 28 mil geometry, this resolution produced a mesh of approximately 25 million cells.
    The concentration boundary layer thickness, estimated as $d_h/Sh$, is approximately $15\,\mu m$ at the highest $Re$ and $\alpha = 45^\circ$.
        
    The setup involves open flow faces where velocity and scalar variables are subject to pairwise periodic conditions.
    Although the absolute pressure $P$ is not periodic, the dynamic pressure $p$ is, as described in Eq. (\ref{eq:pressureGradient}). 
    At the spacer, a no-slip condition is applied to velocity, and no-gradient conditions are applied to the scalar variables.
    For the top and bottom walls, a no-slip condition is applied to velocity, no-gradient for pressure, and the concentration is fixed to a value of 1. 
    The initial field has the bulk concentration set to zero, resulting in a concentration difference $\Delta s = 1$. 
    The cross flow is established using a prescribed pressure gradient to drive the flow towards a target cross-flow velocity, which is achieved using OpenFOAM's 'meanVelocityForce' feature.

\section{Results}\label{sec:results}

\subsection{Simulation Data}

\begin{figure}[H]
        \centering
        \includegraphics[width=0.65\linewidth]{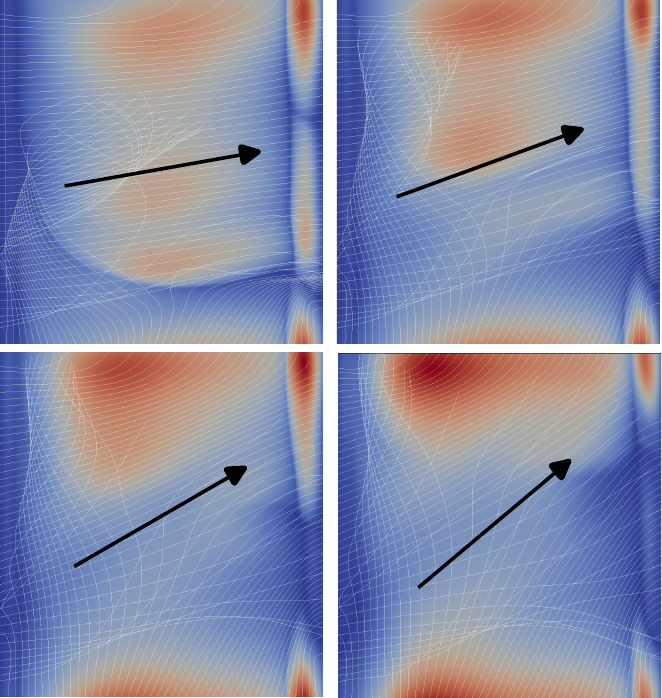}
        \caption{Map of scalar concentration gradient. The colors indicate poor mixing in dark blue areas, $\frac{\partial_n s}{\Delta s}=0$, and strong mass transfer in dark red areas with $\frac{\partial_n s}{\Delta s}=7\cdot 10^4$. The arrows indicate the direction of the pressure drop $\alpha\in \{ 10^\circ,20^\circ,30^\circ,40^\circ \}$}.
        \label{fig:soluteGradMap}
    \end{figure}

    Figure \ref{fig:soluteGradMap} shows how the local wall-normal concentration gradient, i.e. the local mass transport, changes with the flow orientation.
    The blue shades indicate areas of low scalar gradients and poor mixing, while red shades indicate local areas of increased mass transport or low concentration polarization.
    
    The mass transport is quantified by the area-averaged magnitude of the non-dimensional wall-normal scalar gradient, Eq. (\ref{eq:wallNormalScalarGrad}), which is implemented in the OpenFOAM function object 'wallNormalScalarGrad'\footnote{\url{https://bitbucket.org/FynnAschmoneit/wallnormalscalargrad/src/master/wallNormalScalarGrad/}}.
    
    \begin{equation}\label{eq:wallNormalScalarGrad}
        \frac{\partial_n s}{\Delta s} = \frac{1}{\Delta s\, A_{top,bottom}} \sum_{i: \text{top,bottom faces}} | \partial_n s_i | \,  A_i
    \end{equation}
    
    The pressure drop is quantified through the integral shear forces on the surfaces.
    The local reduced shear stress $S_i$, computed with OpenFOAM's incompressible 'wallShearStress' function, is averaged over all wetted boundary faces $A_i$:
    
    \begin{equation}\label{eq:wallShearForce}
        \frac{\tau}{\rho} = \frac{1}{A_{wet}} \sum_{i: \text{wall faces}} | S_i | \,  A_i
    \end{equation}

    Table \ref{tab:raw_data} shows the simulation results for mass transport and shear forces, according to Eqs. (\ref{eq:wallNormalScalarGrad}, \ref{eq:wallShearForce}), against cross flow and flow orientation.
    The kinematic viscosity is set to $\nu = 1 \times 10^{-6} m^2/s$ and the diffusion coefficient of NaCl in water is $D=1.49\times 10^{-9} m^2/s$.
    Orientation values denoted as '–' refer to the spacer-free channel, included as a reference baseline.
    
    \begin{table}[H]
    \begin{center}
    \caption{Simulation results for narrow and wide channel geometries, with feed spacer at different orientations.}     \label{tab:raw_data}
        \begin{tabular}{cccccc} \hline
           \textbf{orientation}  & \textbf{cross flow} & \multicolumn{2}{c}{ \textbf{scalar normal grad.} } & \multicolumn{2}{c}{ \textbf{red. wall shear force} }  \\ 
            $\alpha\, [deg]$  & $U \, [cm \, s^{-1}]$    & \multicolumn{2}{c}{ $\partial_n s / \Delta s \,[10^2\,\,m^{-1}]$ } &  \multicolumn{2}{c}{ $\tau/\rho \, [10^{-6}\,\,m^2 s^{-2}]$ }  \\ \hline
               &      & \textbf{28 mil} & \textbf{34 mil}& \textbf{28 mil} & \textbf{34 mil} \\ \hline
            0  & 5    & 271 & 245 & 635  & 532  \\
            0  & 10   & 382 & 357 & 1588 & 1385  \\
            0  & 15   & 475 & 435 & 2913 & 2572  \\
            0  & 20   & 552 & 503 & 4529 & 4025  \\ \hline
            10  & 5   & 309 & 276 & 693  & 581  \\
            10  & 10  & 427 & 386 & 1714 & 1490  \\
            10  & 15  & 518 & 471 & 3122 & 2752  \\
            10  & 20  & 590 & 537 & 4847 & 4302  \\  \hline
            20  & 5   & 331 & 294 & 745  & 622  \\
            20  & 10  & 450 & 405 & 1823 & 1580  \\
            20  & 15  & 554 & 496 & 3325 & 2917  \\
            20  & 20  & 643 & 571 & 5182 & 4569  \\  \hline
            30  & 5   & 344 & 306 & 773  & 645  \\
            30  & 10  & 473 & 427 & 1883 & 1634  \\
            30  & 15  & 580 & 525 & 3422 & 3006  \\
            30  & 20  & 679 & 611 & 5314 & 4703  \\ \hline
            40  & 5   & 351 & 312 & 786  & 656  \\
            40  & 10  & 475 & 432 & 1896 & 1640  \\
            40  & 15  & 576 & 535 & 3432 & 3008  \\
            40  & 20  & 673 & 631 & 5330 & 4704  \\  \hline
            -  & 5    & 142 & 112 & 45   & 37  \\
            -  & 10   & 181 & 156 & 90   & 75  \\
            -  & 15   & 214 & 181 & 148  & 128  \\
            -  & 20   & 236 & 206 & 219  & 196  \\ \hline
        \end{tabular} 
        \end{center}
    \end{table}

    Qualitatively, the data suggest that the shear forces in the spacer-filled channels are one order of magnitude greater than in the empty channel, whereas the mass transport increases by less than one order of magnitude.
    Both quantities increase with the cross-flow velocity and the orientation angle, while the narrow channel yields the greater shear stress.
    A systematic analysis of these effects is presented in the following subsection.

\subsection{Orientation-Resolved Scaling Relations for Mass Transport and Flow Resistance}\label{sec:Scaling}

    It is proposed to quantify the effect of the spacer orientation as $| \sin(2\alpha) |$, since the geometry requires a symmetry around $\alpha=0$ and a periodicity of $90^\circ$.
    Without loss of generality, only positive orientations $\alpha \in [0^\circ, 90^\circ]$ are considered.
    The proposed scaling relations for the Sherwood number and the friction factor are:
    
    \begin{equation}\label{eq:SH_scaling_def}
        Sh = A_1 \left( \frac{d_h}{L} \right)^\frac{1}{2} \Bigl( 1 + A_2 \, \sin(2\alpha) \Bigr)  Re^\frac{1}{2}
    \end{equation}

    \begin{equation}\label{eq:f_scaling_def}
        f = B_1 \Bigl( 1 + B_2 \, \sin(2\alpha) \Bigr)  Re^{-\frac{1}{2}}
    \end{equation}

\begin{figure}[H] 
        \centering
            \includegraphics[width=0.5\linewidth]{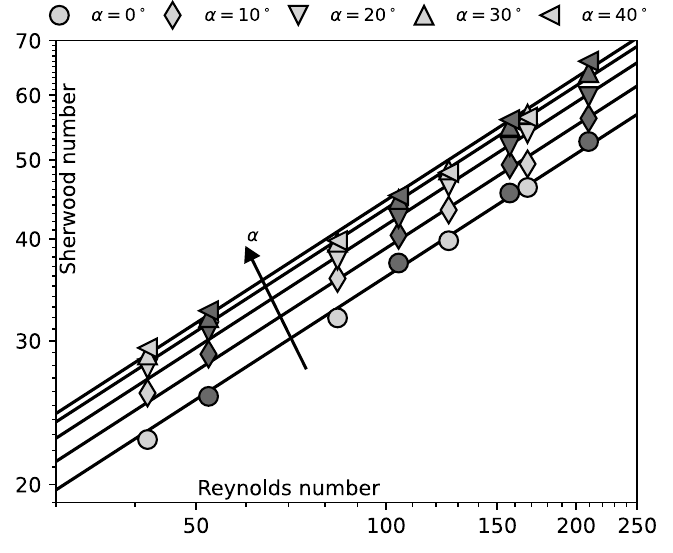} 
        \caption{Mass transport against cross flow and spacer orientation. The bright and dark data points refer to the 28 mil and the 34 mil spacer, respectively. The solid lines show the single fit for the cumulated data sets, according to Eq. (\ref{eq:SH_scaling_result}). }
        \label{fig:SH_fits}
    \end{figure}

\begin{figure}[H] 
        \centering
            \includegraphics[width=0.5\linewidth]{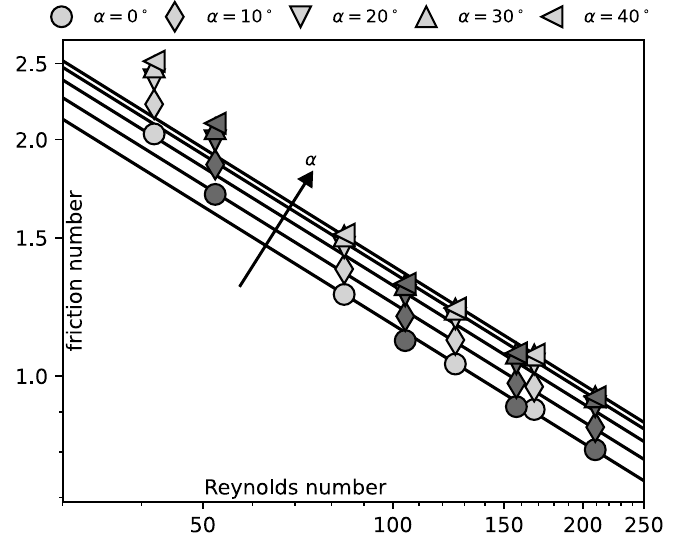} 
        \caption{Flow resistance against cross flow and spacer orientation. The bright and dark data points refer to the 28 mil and the 34 mil spacer, respectively. The solid lines show the single fit for the cumulated data sets, according to Eq. (\ref{eq:f_scaling_result}). The low-Reynolds number values were excluded in the data fit.}
        \label{fig:f_fits}
    \end{figure}

    For the purpose of greatest accuracy and reproducibility, the coefficients $A_{1,2}$ and $B_{1,2}$ are found by an analytic least-square fit of Eqs. (\ref{eq:SH_scaling_def}) and (\ref{eq:f_scaling_def}) to the concentration gradient, the shear stress and the cross-flow data in Tab. \ref{tab:raw_data}, for each spacer individually and for both spacers cumulatively.
    The analytic least-square fitting, the linearization of the scaling relations and the construction of the data vectors are given in \ref{app:leastSquare}.
    Table \ref{tab:fit_coeff_SH_f} lists the resulting coefficients and the goodness-of-fit $R^2$, Eq. (\ref{eq:R2}), for both, the Sherwood and the friction number scaling.
    Figures \ref{fig:SH_fits} and \ref{fig:f_fits} shows the fitted scaling relations together with the according data, for the mass transport and the flow resistance, respectively. 
    As seen in Fig. \ref{fig:f_fits}, the friction factors at the lowest cross-flow velocity, $U = 5\,cm\,s^{-1}$, follow a different scaling behavior and are excluded from the fits.
    All data points in each figure are fitted simultaneously, hence the distinct lines in one figure result from one single fit.
    
    \begin{table}[H]
        \begin{center}
        \caption{Least-square coefficients and goodness-of-fit of Sherwood number scaling (\ref{eq:SH_scaling_def}) and friction number scaling (\ref{eq:f_scaling_def}).}     \label{tab:fit_coeff_SH_f}
        \begin{tabular}{ccccccc}
                \hline
                \textbf{spacer} & \multicolumn{3}{c}{\textbf{Mass transport}} & \multicolumn{3}{c}{\textbf{Flow resistance}}\\
                 & $A_1$  & $A_2$ & $R_A^2\,\, [\%]$ & $B_1$  & $B_2$ & $R_B^2\,\, [\%]$ \\
                \hline
                28 mil  & 6.542 & 0.244 & 99.6 & 11.716 & 0.194 & 99.6 \\
                34 mil  & 6.743 & 0.244 & 99.8 & 11.522 & 0.185 & 99.4\\
                cumulated  & 6.647 & 0.244 & 99.7 & 11.630 & 0.190 & 99.2 \\
                \hline
            \end{tabular}
        \end{center}
    \end{table}

\subsubsection*{Cumulated Scaling Relations}

    With the cumulative coefficients in Tab. \ref{tab:fit_coeff_SH_f} the scaling relations for both spacers are:

    \begin{align}
        \label{eq:SH_scaling_result}
        &Sh = 6.65 (1 + 0.244 \sin(2 \alpha) ) \left( \frac{d_h}{L} \right)^\frac{1}{2}  Re^\frac{1}{2}  \\
        \label{eq:f_scaling_result}
        &f = 11.63 ( 1 + 0.19 \sin(2 \alpha) ) Re^{-\frac{1}{2}}
    \end{align}

\begin{figure}[H] 
        \centering
        \subfigure[]{
            \includegraphics[width=0.45\linewidth]{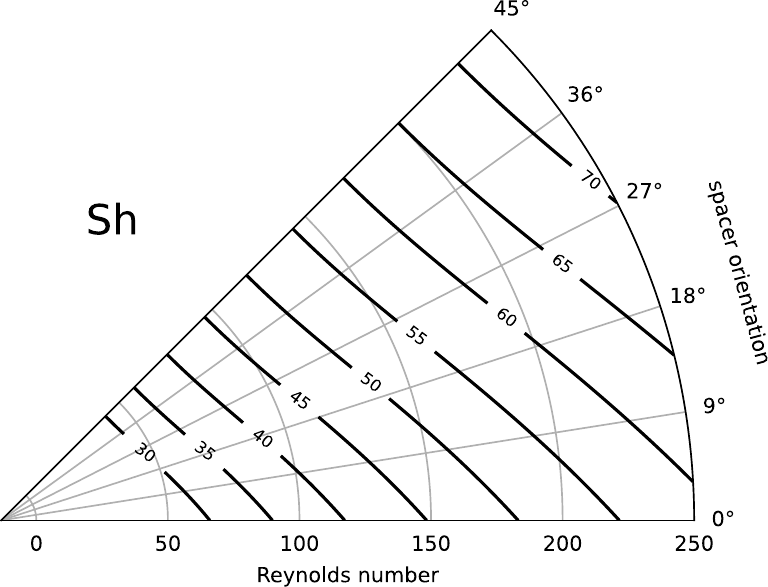} }
        \subfigure[]{            
            \includegraphics[width=0.45\linewidth]{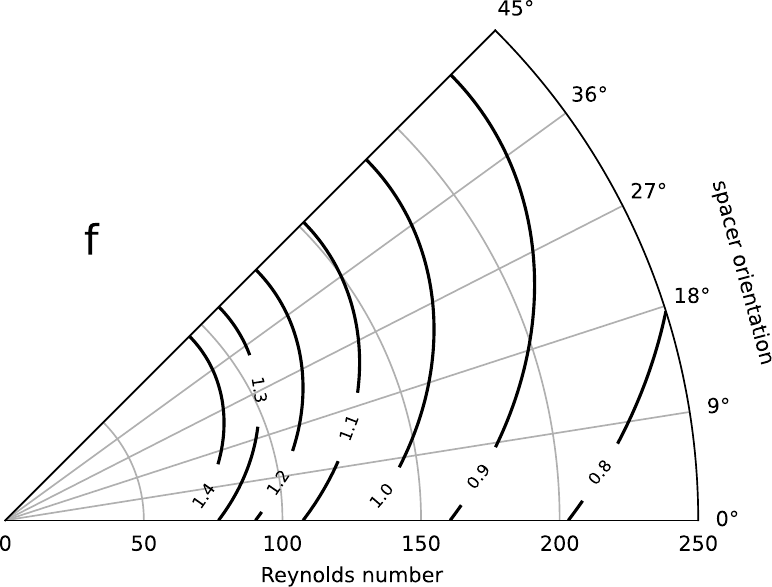} }
        \caption{Contour plots of $Sh$ (a) and $f$ (b) in cylindrical coordinates of $Re$ and $\alpha$.}
        \label{fig:pieplot}
    \end{figure}

    The scaling relations for $Sh$ and $f$ in Eqs. (\ref{eq:SH_scaling_result}) and (\ref{eq:f_scaling_result}) accurately capture the effect of $\alpha$ and $Re$ for both spacers simultaneously.
    They constitute the central result of this article and form the basis for the remaining analysis.
    Figure \ref{fig:pieplot} displays their contours in a cylindrical coordinate plot, with the cylindrical distance displaying $Re$ and the azimuth angle displaying $\alpha$.
    The contours of $Sh$ and $f$ cross those of constant $Re$, which emphasizes the significant effect of $\alpha$ on both quantities.
    
\subsection{Implications for Reverse Osmosis Performance}\label{sec:performaceAnalysis}

    With the scaling relations established, the question becomes how the orientation affects the water flux $j_w$ and the pressure drop $\partial_x p$ in practice, given that a $45^\circ$ orientation increases $Sh$ by 24\% and $f$ by 19\%.
    At constant cross flow and increasing $\alpha$, the trade-off between increasing $Sh$ and $f$ is analyzed through the derived scaling relations and the Algebraic Water Flux Framework (AWFF) \citep{tiraferri_standardizing_2023} in the following.
    Coupling CFD-derived correlations for the Sherwood number and the friction factor with module-scale models has been recommended for assessing spacer performance at full scale \citep{liang_role_2025}.

    The pressure drop $\partial_x p$ follows from the Darcy-Weisbach equation (\ref{eq:DarcyWeisbach}), which defines the flow resistance as the application-specific friction factor times the volumetric kinetic fluid energy, per representative length-scale.

    \begin{equation}\label{eq:DarcyWeisbach}
        \frac{\partial p}{\partial x} = \frac{f}{d_h} \frac{\rho U^2}{2}
    \end{equation}
    
    Deriving the water flux $j_w$ from bulk flow variables is complicated by concentration polarization (CP), which enters through an exponential term in the water flux Eq. (\ref{eq:waterFluxTraditional}), in which $k_d$ denotes the mass transport coefficient, $A$ the water permeability coefficient, $p_f$ the feed pressure and $\pi_f$ the osmotic pressure in the feed.
    Since $j_w$ appears on both sides of Eq. (\ref{eq:waterFluxTraditional}), no closed-form solution exists and iterative numerical solvers are required.

    \begin{equation}\label{eq:waterFluxTraditional}
        j_w = A \left( p_f - \pi_f \exp{ \left( \frac{j_w}{k_d} \right) }  \right)
    \end{equation}
        
    The AWFF circumvents this problem by a closed-form approximation of $j_w$ in terms of the reduced pressure $P$ and the reduced mass transport $K$, in Eq. (\ref{eq:J_algebraicWaterFlux}):
        \begin{equation}\label{eq:J_algebraicWaterFlux}
        \frac{j_w}{A(p_f - R \pi_f)} = 1 - \frac{1}{1+K} - \frac{PK}{2(1+K)^3} \underset{K \gg 1}{\approx} \frac{K}{1+K}
    \end{equation}
    Here, $P$ is the ratio of the feed pressure $p_f$ to the osmotic pressure $\pi_f$ minus the observed rejection $R$, while $K$ is the ratio of the mass transport coefficient $k_d = Sh\,D/d_h$ to the theoretical counter flow $A \pi_f$:
    
    \begin{equation}\label{eq:algWaterFlux_nonDimDef}
    \begin{split}
        P = \frac{p_f}{\pi_f} - R, \quad K = \frac{k_d}{A \pi_f}, 
    \end{split}
    \end{equation}

    A constant $R$ is adopted here rather than resolving its downstream variation along the membrane, since the resulting closed-form estimate for $j_w$ is intended for rapid, design-stage screening of orientation effects, for which the tractability of Eq. (\ref{eq:J_algebraicWaterFlux}) outweighs the marginal accuracy gain from a spatially resolved rejection profile.
    The relative flux gain from orientation is insensitive to this choice: for $R$ between 0.98 and 0.999 for SWRO and between 0.95 and 0.99 for BWRO, the absolute water flux changes by about 1\%, whereas the relative flux gain between $0^\circ$ and $45^\circ$ changes by less than 0.01 percentage points.

   \begin{figure}[H]
       \centering
        \begin{tikzpicture}[node distance=1.4cm]
        
            \node[block]                    (Re)   {$Re$};
            \node[block, below=of Re]       (Shf)  {$Sh,\; f$};
            \node[block, below=of Shf]      (PK)   {$P,\; K$};
        
            \node[input, left=2cm of Re]    (in1)  {Fluid, flow \& spacer\\$\rho, \; \mu, \; D, \; U,\; d_h,\;  \alpha$};
            \node[input, left=2cm of PK]    (in2)  {Membrane \\$p_f,\; \pi_f,\; R,\; A$};
        
            \node[output, right=2cm of Shf] (dp)   {$\frac{\partial p}{\partial x}$};
            \node[output, right=2cm of PK]  (jwout){$j_w$};
        
            
            \draw[arrow] ([xshift=-0.9cm]Re.south)  --node[right,align=center]{Scaling rel.\\ eqs. (\ref{eq:SH_scaling_result}),(\ref{eq:f_scaling_result})} ([xshift=-0.9cm]Shf.north);
            \draw[arrow] ([xshift=-0.9cm]Shf.south) --node[right,align=center]{AWFF \\eq. (\ref{eq:algWaterFlux_nonDimDef})} ([xshift=-0.9cm]PK.north);
            
            \draw[arrow] (in1) -- (Re);
            \draw[arrow] (in2) -- (PK);
        
            \draw[arrow] ([yshift=-0.2cm]Shf.east) --node[above,align=center]{D-W (\ref{eq:DarcyWeisbach})} ([yshift=-0.2cm]dp.west);
            \draw[arrow] ([yshift=-0.2cm]PK.east) --node[above,align=center]{eq. (\ref{eq:J_algebraicWaterFlux})} ([yshift=-0.2cm]jwout.west);

            \begin{pgfonlayer}{background}
        
                \node[colbox, fit=(in1)(in2), label={[anchor=south, font=\small\bfseries, align=center]north: Bulk Flow and \\ Membrane Variables}]
                    (leftbox) {};
        
                \node[colbox, fit=(Re)(Shf)(PK), label={[anchor=south, font=\small\bfseries, align=center]north:Nondim. Flow- and \\Process Variables}]
                    (centerbox) {};
        
                \node[colbox, fit=(dp)(jwout), label={[anchor=south, font=\small\bfseries, align=center]north:Target Quantities}]
                    (rightbox) {};
        
            \end{pgfonlayer}
        
        \end{tikzpicture}
       \caption{Flow diagram for analyzing process performance, with scaling relations and AWFF.}
       \label{fig:tikzDiagram}
   \end{figure}

    Figure \ref{fig:tikzDiagram} illustrates how $\partial_x p$ and $j_w$ are evaluated from the bulk variables for the flow and the membrane.
    The workflow is applied to two representative cases, namely seawater (SWRO) and brackish water (BWRO), comparing spacer orientations $\alpha=0^\circ$ and $\alpha=45^\circ$ at $U = 0.2$ m/s. 
    The respective variables for the bulk flow and the membrane, i.e. the input variables, are summarized in Tab. \ref{tab:membrane_processes_SW}, together with the derived target quantities $\partial_x p$ and $j_w$ in the bottom rows.

    \begin{table}[H]
        \centering
        \begin{tabular}{lcccc}
        \hline
        \textbf{Bulk Variables } &  \dcl{\textbf{ Seawater}} & \dcl{\textbf{Brackish Water}} \\
        \hline
        $d_h\hfill[\mu m]$           &  \dcl{838}      & \dcl{1046} \\
        $A \; \; \hfill [L \;m^{-2} h^{-1} bar^{-1}$] & \dcl{1.1}        & \dcl{4} \\
        Rejection R        & \dcl{0.997}     & \dcl{0.980} \\
        $p_f \hfill [bar]$         & \dcl{70} & \dcl{20} \\
        $\pi_f \hfill [bar]$       & \dcl{27} & \dcl{4} \\
        \hline
        \textbf{Target Quantities} & $\alpha=0^\circ$ & $\alpha=45^\circ$ & $\alpha=0^\circ$ & $\alpha=45^\circ$ \\
        \hline
        $\partial_x p \hfill [kPa\; m^{-1}]$ & 21.44 & 25.51  & 15.37 & 18.30 \\
        $j_w \hfill [L \; m^{-2} h^{-1}]$ & 42.69 & 43.58 & 60.17 & 61.01 \\
        \hline
        \end{tabular}
        \caption{Bulk Variables and Target Quantities for quantitative performance assessment.}
        \label{tab:membrane_processes_SW}
    \end{table}

    In both applications, $\alpha=45^\circ$ increases both $\partial_x p$ and $j_w$, consistent with the scaling relations for $f$ and $Sh$.
    The workflow in Fig. \ref{fig:tikzDiagram} yields the following quantitative trade-off:
    The relative increase in water flux, $(j_{w\;(45^\circ)} - j_{w\;(0^\circ)})/ j_{w\;(0^\circ)}$, is marginal in both filtration applications (2.09\% for SWRO and 1.39\% for BWRO).
    The relative increase in pressure drop is 19\%, both for SWRO and for BWRO, which directly reflects the scaling coefficient $B_2$ in Tab. \ref{tab:fit_coeff_SH_f}.
    Therefore, when optimizing a membrane module for a particular filtration application, it must be carefully considered whether the marginal increase in water flux outweighs the significant increase in pressure drop.

\section{Discussion}\label{sec:discussion}

\subsection*{Scaling exponents and effect of orientation}

    The scaling exponents agree with the ranges reported in the literature: $Sh$ follows $Re^{1/2}$, which lies within the range for laminar and transitional flows \citep{da_costa_spacer_1994, koutsou_numerical_2009, gurreri_flow_2016, fimbres-weihs_numerical_2007, shi_performance_2015}.
    When the exponent is fitted instead of prescribed, the data in Tab. \ref{tab:raw_data} yield $n = 0.498 \pm 0.006$, which supports the prescribed value.
    The friction factor follows $Re^{-1/2}$ for $Re>80$, at the upper end of the values reported for transitional flows \citep{shi_performance_2015, farkova_pressure_1991, mokhtar_cfd_2021}, but becomes steeper towards lower $Re$ and approaches the laminar scaling $m=1$ \citep{da_costa_spacer_1994, mokhtar_cfd_2021}.
    A single power law therefore does not describe $f$ over the full $Re$ range, in line with \citet{mokhtar_cfd_2021}.

    At $\alpha = 0^\circ$, the filaments of one layer are aligned with the flow and hardly obstruct it, whereas with increasing $\alpha$ both filament layers obstruct and deflect the flow.
    This enlarges the regions of high wall-normal concentration gradient (Fig. \ref{fig:soluteGradMap}) and increases the wall shear stress (Tab. \ref{tab:raw_data}).
    Since the relative increase from $0^\circ$ to $45^\circ$ is larger for $Sh$ (24\%) than for $f$ (19\%), orientation enhances mass transport more than it increases the friction factor at constant cross-flow velocity.

\subsection*{Comparison with previous studies}

    The present results agree with previous studies in identifying $45^\circ$ as the orientation with the best mixing \citep{gurreri_flow_2016, albeirutty_experimental_2018, haidari_determining_2019, mokhtar_cfd_2021}.
    \citet{ponzio_experimental_2017} found the heat transfer advantage of $45^\circ$ to decrease with increasing $Re$, over a range extending into unsteady flow.
    In the steady range considered here, the relative gain in $Sh$ is approximately independent of $Re$, as expressed by the separable form of Eq. (\ref{eq:SH_scaling_result}).
    For the pressure drop, the literature is not consistent: for woven spacers, the lowest friction coefficient was found at $45^\circ$ \citep{mokhtar_cfd_2021}, whereas higher pressure losses at $45^\circ$ than at $0^\circ$ were measured for commercial spacers \citep{haidari_determining_2019}.
    The overlapped spacers studied here show an increase of $f$ by 19\% from $0^\circ$ to $45^\circ$, so the orientation effect on the flow resistance depends on the spacer type and results for woven spacers cannot be transferred to overlapped spacers.
    While previous studies report the orientation effect at selected angles, the present relations express the orientation dependence of both $Sh$ and $f$ as a continuous function of $\alpha$ that is separable from the $Re$ dependence.
    For the two commercial spacer sizes studied, one set of coefficients describes both spacers.
    Whether this holds for other spacer geometries remains to be tested.

\subsection*{Implications for reverse osmosis performance}

    The flux gain from orientation, 2.09\% for SWRO and 1.39\% for BWRO, is small compared to the gain in $Sh$, which follows from Eq. (\ref{eq:J_algebraicWaterFlux}).
    For strong mixing, $K \gg 1$, the water flux is proportional to $K/(1+K)$.
    Since $K \propto k_d \propto Sh$, a rotation from $0^\circ$ to $45^\circ$ increases $K$ by the factor $1+A_2$.
    The relative flux gain therefore decays as $1/K= \frac{A \pi_f}{k_d}$:
    
    \begin{equation*}
        \frac{j_w(\alpha = 45)-j_w(\alpha=0)}{j_w(\alpha=0)} 
        = \frac{ \frac{(1+A_2)K}{1+(1+A_2)K} - \frac{K}{1+K}}{ \frac{K}{1+K} }  = \frac{A_2}{1 + (1+A_2)K}
    \end{equation*}

    For the conditions in Tab. \ref{tab:membrane_processes_SW}, $K \approx 10$ for SWRO and $K \approx 17$ for BWRO, which gives 1.8\% and 1.1\%, close to the values obtained with the full equation.
    The benefit of orientation therefore increases as $K$ decreases, i.e. for more permeable membranes and higher osmotic pressures, where concentration polarization is more severe \citep{sreedhar_evolution_2023, liang_role_2025}.
    The pressure drop increases by 19\% at constant cross-flow velocity, which corresponds to 4.1~kPa per metre of module length for SWRO at a feed pressure of 70~bar.
    Whether the flux gain outweighs the additional pressure loss requires an energy balance at module scale, which is beyond the present local analysis, in particular since the flux gain of high-performing spacers decreases along the membrane length as the cross-flow velocity declines \citep{liang_role_2025}.

\subsection*{Limitations}

    The scaling relations are derived for two overlapped spacers with circular filaments and nearly identical $d_h/L$ of 0.297 and 0.288, so the exponent of $d_h/L$ cannot be determined from the present data.
    The filament diameter, the intersection depth and the mesh angle are fixed, so the coefficients apply to this spacer family only.
    Since the pressure drop depends nonlinearly on the geometric parameters \citep{johannink_predictive_2015}, an extrapolation to other geometries is not justified.
    All simulations use a single Schmidt number, $Sc \approx 670$ for NaCl in water, so the $Sc$ dependence of $Sh$ is not resolved.
    The flow is assumed steady, which is supported for a commercial spacer at $0.185\,m\,s^{-1}$ \citep{qamar_novel_2021} but has not been verified here for the highest $Re$ and $\alpha$.

\section{Conclusions}

    This study quantifies the effect of the flow attack angle on mass transport and flow resistance in channels filled with two commercial overlapped feed spacers, using high-resolution CFD simulations of periodic spacer cells.
    The scaling coefficients are obtained by an analytic least-square fit for accuracy and reproducibility, with a hydraulic diameter that accounts for the intersection of the filaments.
    The main conclusions are:

\begin{itemize}
    \item Mass transport and flow resistance are described by $Sh = 6.65(1 + 0.244 \sin(2\alpha)) (d_h/L)^{1/2} Re^{1/2}$ and $f = 11.63(1 + 0.19 \sin(2\alpha)) Re^{-1/2}$, in which the effects of $Re$ and $\alpha$ are separable.
    To the authors' knowledge, this is the first continuous relation for the dependence of both $Sh$ and $f$ on the flow attack angle.
    \item One set of coefficients describes both spacer sizes.
    The orientation coefficients are nearly identical for the two spacers, 0.244 for $Sh$ and 0.194 and 0.185 for $f$.
    \item Rotating the spacer from $0^\circ$ to $45^\circ$ increases $Sh$ by 24\% and $f$ by 19\% at constant cross-flow velocity.
    \item $Sh$ follows $Re^{1/2}$ over the full range studied, while $f$ follows $Re^{-1/2}$ for $Re>80$ and becomes steeper towards lower $Re$.
    \item Combined with the algebraic water flux framework, the scaling relations give closed-form estimates of the water flux and the pressure drop.
    The relative flux gain from orientation decays inversely with $K$.
    It amounts to 1\% to 2\% for typical SWRO and BWRO conditions but increases for more permeable membranes.
\end{itemize}

    The relations are valid for overlapped spacers with circular filaments at $Sc \approx 670$, for $Re$ between about 40 and 210 for $Sh$ and between 80 and 210 for $f$.
    Future work should extend them to other spacer geometries and Schmidt numbers and to the module scale, including an energy balance of the orientation trade-off.
    It should also be investigated whether the orientation dependence can be incorporated in an orientation-dependent hydraulic diameter.

\section{Data Availability and Reproducibility Statement}
All data are summarized in Tab. \ref{tab:raw_data}, and the referenced OpenFoam case may be used to reproduce these data.

\appendix
\section{Analytic Least-Square Fitting}\label{app:leastSquare}

    The general methodology is rooted in fitting the linear function $a X + b Y$, with the independent vector variables $X,Y$, and the coefficients $a,b$, against the data encoded in a vector $Z$.
    The least-square solution is found by minimizing the sum of the squared residuals: 
    
    \begin{equation}
    \begin{split}
    \label{eq:chi2}
       \chi^2 &= \sum_{i=1}^{5N} \left(  Z_i - a X_i - b Y_i   \right)^2\\
       &= ( Z - aX - bY ) \cdot ( Z - aX - bY  )
   \end{split}
   \end{equation}

    with:
    \begin{align*}
        &\frac{\partial}{\partial a} \chi^2 = -2 X \cdot ( Z - aX -bY ) \\
        &\frac{\partial}{\partial b} \chi^2 = -2 Y \cdot ( Z - aX -bY ) ,
    \end{align*}

    where the  dot denotes the inner product of the vector variables.
    The least-square solution follows from $\partial_a \chi^2 = 0$ and $\partial_b \chi^2 = 0$, which yields the following linear system for $(a,b)$:


    \begin{equation}    \label{eq:leastSquare_linEqSystem}
    \begin{pmatrix}
    X\cdot X & X\cdot Y \\
    X\cdot Y & Y\cdot Y
    \end{pmatrix} 
    \begin{pmatrix}
    a  \\
    b 
    \end{pmatrix} =
    \begin{pmatrix}
    X\cdot Z  \\
    Y\cdot Z 
    \end{pmatrix} 
    \end{equation}
    
    As a consequence of unknown uncertainties for the measurements in $Z$, the goodness-of-fit is assessed using the mean of the data vector $\bar{Z}$, with  $\sigma^2 = (Z - \bar{Z}) \cdot (Z - \bar{Z}) $, as:

    \begin{equation}
    \label{eq:R2}
        R^2 = 1 - \frac{\chi^2}{ \sigma^2}
    \end{equation}

    In order to apply the least-square fitting in Eqs. (\ref{eq:chi2}-\ref{eq:R2}) to the different data sets in Tab. \ref{tab:raw_data}, the respective scaling functions need to be linearized.

\subsection{Linearization of the Sherwood Number Scaling}\label{app:linSh}

    The scaling relation (\ref{eq:SH_scaling_def}) is first linearized in the form $aX + bY$ and then fitted by Eq. (\ref{eq:leastSquare_linEqSystem}).
    Here, the independent variables $(X,Y)$ are constructed from the independent variables $(Re, \alpha)$.        
    With the rotation variable $Q = \sin(2 \alpha)$, the scaling relation (\ref{eq:SH_scaling_def}) is linearized as:
    \begin{equation*}
        Sh = \underbrace{A_1}_a \underbrace{ \left( \frac{d_h}{L} \right)^\frac{1}{2}  Re^{\frac{1}{2}}}_X + \underbrace{A_1A_2}_b\, \underbrace{\sin(2 \alpha) \left( \frac{d_h}{L} \right)^\frac{1}{2} Re^{\frac{1}{2}}}_Y
    \end{equation*}
    
    Note that similar cross flow velocities in both spacers translate to different Reynolds numbers, as a consequence of the different hydraulic diameters.   
    Let $N \in \{ 4,8 \}$ denote the number of distinct Reynolds numbers for one spacer or both spacers collectively and let $Q_i = \sin(2 \alpha_i)$, $i\in[1,5]$, denote the values of the rotation variable for the angles $\alpha_i$ in Tab. \ref{tab:raw_data}.
    The vector $X$ then consists of five consecutive repetitions of the $N$ distinct Reynolds number terms.
    In $Y$, each block of $N$ consecutive entries of $X$ is multiplied by one of the five values $Q_i$.
    The vector $Z$ contains the corresponding Sherwood numbers from Tab. \ref{tab:raw_data}:
    
    \begin{equation*}
    \begin{split}
        X =& 
        \left( \frac{d_h}{L} \right)^\frac{1}{2} \Bigl(  \underbrace{Re_1^\frac{1}{2}, Re_2^\frac{1}{2}, ..., Re_N^\frac{1}{2}}_\text{1}, \quad 
        \underbrace{Re_1^\frac{1}{2}, Re_2^\frac{1}{2}, ..., Re_N^\frac{1}{2}}_\text{2}, \quad , ..., \quad
        \underbrace{Re_1^\frac{1}{2}, Re_2^\frac{1}{2}, ..., Re_N^\frac{1}{2}}_\text{5} \Bigr)\\
        Y =& \Bigl(
        \underbrace{Q_1 X_1, Q_1 X_2, ..., Q_1 X_N}_\text{1}, \quad 
        \underbrace{Q_2 X_{N+1}, Q_2 X_{N+2}, ..., Q_2 X_{2N}}_\text{2}, \quad , ..., \\
        & \quad , ..., \quad 
        \underbrace{Q_5 X_{4 N+1}, Q_5 X_{4 N+1}, ..., Q_5 X_{5N}}_\text{5}     \Bigl) \\
        Z =& \Bigl(     Sh_1, Sh_2, ..., Sh_{5N}    \Bigr)
    \end{split}
    \end{equation*}

    The coefficients $(A_1,A_2)$ in Eq. (\ref{eq:SH_scaling_def}) follow from the least-square coefficients $(a,b)$ as $A_1 = a$ and $A_2 = b/a$.

\subsection{Linearization of the Friction Factor Scaling}\label{app:linf}

    Similarly to the case of the Sherwood number, the scaling relation for the friction factor (\ref{eq:f_scaling_def}) is linearized with the independent variables $(X,Y)$:
    \begin{equation*}
        f = \underbrace{B_1}_a \underbrace{Re^{-\frac{1}{2}}}_X + \underbrace{B_1 B_2}_b\, \underbrace{Q Re^{-\frac{1}{2}}}_Y,
    \end{equation*}

    The friction factor data at the lowest cross-flow velocity are excluded from the fit, see Section \ref{sec:Scaling}.
    The number of distinct Reynolds numbers is therefore $N=\left\{ 3,6 \right\}$, depending if only one spacer, or both spacers are considered.
    
    \begin{equation*}
    \begin{split}
        X =&  \Bigl(
         \underbrace{Re_1^{-\frac{1}{2}}, Re_2^{-\frac{1}{2}}, ..., Re_N^{-\frac{1}{2}}}_\text{1}, \quad 
        \underbrace{Re_1^{-\frac{1}{2}}, Re_2^{-\frac{1}{2}}, ..., Re_N^{-\frac{1}{2}}}_\text{2}, \quad , ..., \quad
        \underbrace{Re_1^{-\frac{1}{2}}, Re_2^{-\frac{1}{2}}, ..., Re_N^{-\frac{1}{2}}}_\text{M} \Bigr) \\
        Y =& \Bigl(
        \underbrace{Q_1 X_1, Q_1 X_2, ..., Q_1 X_N}_\text{1}, \quad 
        \underbrace{Q_2 X_{N+1}, Q_2 X_{N+2}, ..., Q_2 X_{2N}}_\text{2}, \quad , ..., \\
        & \quad , ..., \quad 
        \underbrace{Q_M X_{(M-1)N+1}, Q_M X_{(M-1)N+1}, ..., Q_M X_{MN}}_\text{M}     \Bigl) \\
        Z =& \Bigl( f_1, f_2, ..., f_{MN} \Bigr)
    \end{split}
    \end{equation*}

    Solving the least-square problem (\ref{eq:leastSquare_linEqSystem}) yields the scaling coefficients $B_1 = a$ and $B_2 = b/a$.

\bibliography{autoreferences}

@article{luo_supercomputing_2023,
  author  = {Luo, Jiu and Li, Mingheng and Hoek, Eric M. V. and Heng, Yi},
  title   = {Supercomputing and machine learning-aided optimal design of high permeability seawater reverse osmosis membrane systems},
  journal = {Science Bulletin}, volume = {68}, pages = {397--407}, year = {2023},
  doi     = {10.1016/j.scib.2023.01.039}}

@article{chen_rapid_2024,
  author  = {Chen, Ke and Luo, Jiu and Chen, Junzhi and Lu, Yutong and others and Heng, Yi},
  title   = {A rapid-convergent particle swarm optimization approach for multiscale design of high-permeance seawater reverse osmosis systems},
  journal = {Communications Engineering}, volume = {3}, pages = {149}, year = {2024},
  doi     = {10.1038/s44172-024-00289-y}}

@article{jeong_cfd_2026,
  author  = {Jeong, Yu-hyeok and Gu, Boram},
  title   = {A {CFD}-Informed Surrogate Modelling Framework Linking Spacer-Scale Simulations to Process-Scale Reverse Osmosis Modelling},
  journal = {Korean Journal of Chemical Engineering}, volume = {43}, number = {13}, pages = {3535--3555}, year = {2026},
  doi     = {10.1007/s11814-026-00782-4}}

@article{celik_procedure_2008,
  author  = {Celik, I. B. and Ghia, U. and Roache, P. J. and Freitas, C. J. and Coleman, H. and Raad, P. E.},
  title   = {Procedure for estimation and reporting of uncertainty due to discretization in {CFD} applications},
  journal = {Journal of Fluids Engineering}, volume = {130}, number = {7}, pages = {078001}, year = {2008},
  doi     = {10.1115/1.2960953}}

@article{da_costa_spacer_1994,
    title = {Spacer characterization and pressure drop modelling in spacer-filled channels for ultrafiltration},
    volume = {87},
    issn = {0376-7388},
    url = {https://www.sciencedirect.com/science/article/pii/0376738893E0076P},
    doi = {10.1016/0376-7388(93)E0076-P},
    language = {en},
    number = {1},
    urldate = {2021-10-21},
    journal = {Journal of Membrane Science},
    author = {Da Costa, A. R. and Fane, A. G. and Wiley, D. E.},
    month = feb,
    year = {1994},
    pages = {79--98},
}

@article{koutsou_numerical_2009,
    title = {A numerical and experimental study of mass transfer in spacer-filled channels: {Effects} of spacer geometrical characteristics and {Schmidt} number},
    volume = {326},
    issn = {0376-7388},
    shorttitle = {A numerical and experimental study of mass transfer in spacer-filled channels},
    url = {https://www.sciencedirect.com/science/article/pii/S0376738808008946},
    doi = {10.1016/j.memsci.2008.10.007},
    number = {1},
    urldate = {2026-01-02},
    journal = {Journal of Membrane Science},
    author = {Koutsou, C. P. and Yiantsios, S. G. and Karabelas, A. J.},
    month = jan,
    year = {2009},
    pages = {234--251},
}

@article{gurreri_flow_2016,
    title = {Flow and mass transfer in spacer-filled channels for reverse electrodialysis: a {CFD} parametrical study},
    volume = {497},
    copyright = {https://www.elsevier.com/tdm/userlicense/1.0/},
    issn = {0376-7388},
    shorttitle = {Flow and mass transfer in spacer-filled channels for reverse electrodialysis},
    url = {https://linkinghub.elsevier.com/retrieve/pii/S0376738815301708},
    doi = {10.1016/j.memsci.2015.09.006},
    language = {en},
    urldate = {2025-07-08},
    journal = {Journal of Membrane Science},
    publisher = {Elsevier BV},
    author = {Gurreri, L. and Tamburini, A. and Cipollina, A. and Micale, G. and Ciofalo, M.},
    month = jan,
    year = {2016},
    pages = {300--317},
}

@article{fimbres-weihs_numerical_2007,
    title = {Numerical study of mass transfer in three-dimensional spacer-filled narrow channels with steady flow},
    volume = {306},
    issn = {03767388},
    url = {https://linkinghub.elsevier.com/retrieve/pii/S037673880700614X},
    doi = {10.1016/j.memsci.2007.08.043},
    language = {en},
    number = {1-2},
    urldate = {2021-09-16},
    journal = {Journal of Membrane Science},
    author = {Fimbres-Weihs, G.A. and Wiley, D.E.},
    month = dec,
    year = {2007},
    pages = {228--243},
}

@article{shi_performance_2015,
    title = {Performance of spiral-wound membrane modules in organic solvent nanofiltration – {Fluid} dynamics and mass transfer characteristics},
    volume = {494},
    issn = {03767388},
    url = {https://linkinghub.elsevier.com/retrieve/pii/S0376738815300727},
    doi = {10.1016/j.memsci.2015.07.044},
    language = {en},
    urldate = {2021-09-13},
    journal = {Journal of Membrane Science},
    author = {Shi, Binchu and Marchetti, Patrizia and Peshev, Dimitar and Zhang, Shengfu and Livingston, Andrew G.},
    month = nov,
    year = {2015},
    pages = {8--24},
}

@article{farkova_pressure_1991,
    title = {The pressure drop in membrane module with spacers},
    volume = {64},
    issn = {0376-7388},
    url = {https://www.sciencedirect.com/science/article/pii/037673889180081G},
    doi = {10.1016/0376-7388(91)80081-G},
    language = {en},
    number = {1},
    urldate = {2021-09-30},
    journal = {Journal of Membrane Science},
    author = {Fárková, Jana},
    month = nov,
    year = {1991},
    pages = {103--111},
}

@article{mokhtar_cfd_2021,
    title = {{CFD} prediction of flow, heat and mass transfer in woven spacer-filled channels for membrane processes},
    volume = {173},
    issn = {0017-9310},
    url = {https://www.sciencedirect.com/science/article/pii/S0017931021003495},
    doi = {10.1016/j.ijheatmasstransfer.2021.121246},
    urldate = {2025-02-10},
    journal = {International Journal of Heat and Mass Transfer},
    author = {Mokhtar, Imen El and Gurreri, Luigi and Tamburini, Alessandro and Cipollina, Andrea and Ciofalo, Michele and Bouguecha, Salah al Taher and Micale, Giorgio},
    month = jul,
    year = {2021},
    pages = {121246},
}

@article{haidari_determining_2019,
    title = {Determining effects of spacer orientations on channel hydraulic conditions using {PIV}},
    volume = {31},
    issn = {22147144},
    url = {https://linkinghub.elsevier.com/retrieve/pii/S2214714418303696},
    doi = {10.1016/j.jwpe.2019.100820},
    language = {en},
    urldate = {2021-09-13},
    journal = {Journal of Water Process Engineering},
    author = {Haidari, A.H. and Heijman, S.G.J. and Uijttewaal, W.S.J. and van der Meer, W.G.J.},
    month = oct,
    year = {2019},
    pages = {100820},
}

@article{tiraferri_standardizing_2023,
    title = {Standardizing practices and flux predictions in membrane science via simplified equations and membrane characterization},
    volume = {6},
    copyright = {2023 Springer Nature Limited},
    issn = {2059-7037},
    url = {https://www.nature.com/articles/s41545-023-00270-w},
    doi = {10.1038/s41545-023-00270-w},
    language = {en},
    number = {1},
    urldate = {2023-10-27},
    journal = {npj Clean Water},
    publisher = {Nature Publishing Group},
    author = {Tiraferri, Alberto and Malaguti, Marco and Mohamed, Madina and Giagnorio, Mattia and Aschmoneit, Fynn Jerome},
    month = aug,
    year = {2023},
    note = {Number: 1},
    pages = {1--12},
}

@article{aschmoneit_omsd_2020,
    title = {{OMSD} – {An} open membrane system design tool},
    volume = {233},
    issn = {1383-5866},
    url = {https://www.sciencedirect.com/science/article/pii/S1383586619326097},
    doi = {10.1016/j.seppur.2019.115975},
    language = {en},
    urldate = {2021-04-28},
    journal = {Separation and Purification Technology},
    author = {Aschmoneit, F. J. and Hélix-Nielsen, C.},
    month = feb,
    year = {2020},
    pages = {115975},
}

@misc{aschmoneit_volume_2025,
    title = {Volume and {Surface} {Area} of two {Orthogonal}, {Partially} {Intersecting} {Cylinders}: {A} {Generalization} of the {Steinmetz} {Solid}},
    shorttitle = {Volume and {Surface} {Area} of two {Orthogonal}, {Partially} {Intersecting} {Cylinders}},
    url = {http://arxiv.org/abs/2512.22555},
    doi = {10.48550/arXiv.2512.22555},
    urldate = {2025-12-30},
    publisher = {arXiv},
    author = {Aschmoneit, Fynn Jerome and Cockx, Bastiaan},
    month = dec,
    year = {2025},
    note = {arXiv:2512.22555 [cs]},
}

@article{patankar_fully_1977,
    title = {Fully {Developed} {Flow} and {Heat} {Transfer} in {Ducts} {Having} {Streamwise}-{Periodic} {Variations} of {Cross}-{Sectional} {Area}},
    volume = {99},
    issn = {0022-1481},
    url = {https://doi.org/10.1115/1.3450666},
    doi = {10.1115/1.3450666},
    number = {2},
    urldate = {2021-08-09},
    journal = {Journal of Heat Transfer},
    author = {Patankar, S. V. and Liu, C. H. and Sparrow, E. M.},
    month = may,
    year = {1977},
    pages = {180--186},
}

@article{guan_hydrodynamic_2023,
    title = {Hydrodynamic effects of non-uniform feed spacer structures on energy loss and mass transfer in spiral wound module},
    volume = {673},
    issn = {0376-7388},
    url = {https://www.sciencedirect.com/science/article/pii/S0376738823001357},
    doi = {10.1016/j.memsci.2023.121479},
    urldate = {2026-08-11},
    journal = {Journal of Membrane Science},
    author = {Guan, Hui and Lin, Peifeng and Yu, Sanchuan and Hu, Xiao and Li, Xiaojun and Zhu, Zuchao},
    month = may,
    year = {2023},
    pages = {121479},
}

@article{hu_enhanced_2025,
    title = {Enhanced mass transfer and energy efficiency by a biomimetic feed spacer in reverse osmosis membrane modules},
    volume = {713},
    issn = {0376-7388},
    url = {https://www.sciencedirect.com/science/article/pii/S0376738824008846},
    doi = {10.1016/j.memsci.2024.123290},
    urldate = {2026-08-11},
    journal = {Journal of Membrane Science},
    author = {Hu, Xiao and Li, Junjie and Yu, Sanchuan and Zhu, Zuchao and Lin, Peifeng and Li, Xiaojun},
    month = jan,
    year = {2025},
    pages = {123290},
}

@article{sreedhar_evolution_2023,
    title = {The evolution of feed spacer role in membrane applications for desalination and water treatment: {A} critical review and future perspective},
    volume = {554},
    issn = {0011-9164},
    shorttitle = {The evolution of feed spacer role in membrane applications for desalination and water treatment},
    url = {https://www.sciencedirect.com/science/article/pii/S0011916423001376},
    doi = {10.1016/j.desal.2023.116505},
    urldate = {2026-08-11},
    journal = {Desalination},
    author = {Sreedhar, Nurshaun and Thomas, Navya and Ghaffour, Noreddine and Arafat, Hassan A.},
    month = may,
    year = {2023},
    pages = {116505},
}

@article{liang_role_2025,
    title = {Role of spacers in osmotic membrane desalination: {Advances}, challenges, practical and artificial intelligence-driven solutions},
    volume = {201},
    issn = {0957-5820},
    shorttitle = {Role of spacers in osmotic membrane desalination},
    url = {https://www.sciencedirect.com/science/article/pii/S0957582025008547},
    doi = {10.1016/j.psep.2025.107587},
    urldate = {2026-08-11},
    journal = {Process Safety and Environmental Protection},
    author = {Liang, Y. Y.},
    month = sep,
    year = {2025},
    pages = {107587},
}

@article{qamar_novel_2021,
    title = {Novel hole-pillar spacer design for improved hydrodynamics and biofouling mitigation in membrane filtration},
    volume = {11},
    copyright = {2021 The Author(s)},
    issn = {2045-2322},
    url = {https://www.nature.com/articles/s41598-021-86459-w},
    doi = {10.1038/s41598-021-86459-w},
    language = {en},
    number = {1},
    urldate = {2026-08-11},
    journal = {Scientific Reports},
    publisher = {Nature Publishing Group},
    author = {Qamar, Adnan and Kerdi, Sarah and Ali, Syed Muztuza and Shon, Ho Kyong and Vrouwenvelder, Johannes S. and Ghaffour, Noreddine},
    month = mar,
    year = {2021},
    pages = {6979},
}

@article{al-obaidi_optimizing_2024,
    title = {Optimizing {Reverse} {Osmosis} {Feed} {Spacer} {Design} for {Enhanced} {Dimethylphenol} {Removal} from {Wastewater}: {A} {Study} of {Hydrodynamics} and {Performance} {Indicators}},
    volume = {16},
    copyright = {http://creativecommons.org/licenses/by/3.0/},
    issn = {2073-4441},
    shorttitle = {Optimizing {Reverse} {Osmosis} {Feed} {Spacer} {Design} for {Enhanced} {Dimethylphenol} {Removal} from {Wastewater}},
    url = {https://www.mdpi.com/2073-4441/16/6/895},
    doi = {10.3390/w16060895},
    language = {en},
    number = {6},
    urldate = {2026-08-11},
    journal = {Water},
    publisher = {Multidisciplinary Digital Publishing Institute},
    author = {Al-Obaidi, Mudhar A. and Rashid, Farhan Lafta and Ameen, Arman and Kadhom, Mohammed and Mujtaba, Iqbal M.},
    month = jan,
    year = {2024},
    pages = {895},
}

@article{johannink_predictive_2015,
    title = {Predictive pressure drop models for membrane channels with non-woven and woven spacers},
    volume = {376},
    issn = {0011-9164},
    url = {https://www.sciencedirect.com/science/article/pii/S0011916415300321},
    doi = {10.1016/j.desal.2015.07.024},
    urldate = {2026-08-11},
    journal = {Desalination},
    author = {Johannink, Matthias and Masilamani, Kannan and Mhamdi, Adel and Roller, Sabine and Marquardt, Wolfgang},
    month = nov,
    year = {2015},
    pages = {41--54},
}

@article{bucs_effect_2014,
    series = {Special {Issue}: {Desalination} {Using} {Membrane} {Technology}},
    title = {Effect of different commercial feed spacers on biofouling of reverse osmosis membrane systems: {A} numerical study},
    volume = {343},
    issn = {0011-9164},
    shorttitle = {Effect of different commercial feed spacers on biofouling of reverse osmosis membrane systems},
    url = {https://www.sciencedirect.com/science/article/pii/S0011916413005213},
    doi = {10.1016/j.desal.2013.11.007},
    urldate = {2026-08-11},
    journal = {Desalination},
    author = {Bucs, Sz. S. and Radu, A. I. and Lavric, V. and Vrouwenvelder, J. S. and Picioreanu, C.},
    month = jun,
    year = {2014},
    pages = {26--37},
}

@article{ponzio_experimental_2017,
  title = {Experimental and computational investigation of heat transfer in channels filled by woven spacers},
  volume = {104},
  issn = {0017-9310},
  url = {https://www.sciencedirect.com/science/article/pii/S0017931016301703},
  doi = {10.1016/j.ijheatmasstransfer.2016.08.023},
  urldate = {2026-10-29},
  journal = {International Journal of Heat and Mass Transfer},
  author = {Ponzio, Francesco Nicolò and Tamburini, Alessandro and Cipollina, Andrea and Micale, Giorgio and Ciofalo, Michele},
  month = jan,
  year = {2017},
  pages = {163--177},
}

@article{albeirutty_experimental_2018,
  title = {An experimental study for the characterization of fluid dynamics and heat transport within the spacer-filled channels of membrane distillation modules},
  volume = {430},
  issn = {0011-9164},
  url = {https://www.sciencedirect.com/science/article/pii/S0011916417319835},
  doi = {10.1016/j.desal.2017.12.043},
  urldate = {2026-10-29},
  journal = {Desalination},
  author = {Albeirutty, M. and Turkmen, N. and Al-Sharif, S. and Bouguecha, S. and Malik, A. and Faruki, O. and Cipollina, A. and Ciofalo, M. and Micale, G.},
  month = mar,
  year = {2018},
  pages = {136--146},
}

@article{toh_review_2020,
  title = {A {Review} of {CFD} {Modelling} and {Performance} {Metrics} for {Osmotic} {Membrane} {Processes}},
  volume = {10},
  copyright = {https://creativecommons.org/licenses/by/3.0/},
  issn = {2077-0375},
  url = {https://www.mdpi.com/2077-0375/10/10/285},
  doi = {10.3390/membranes10100285},
  language = {en},
  number = {10},
  urldate = {2026-10-30},
  journal = {Membranes},
  publisher = {Multidisciplinary Digital Publishing Institute},
  author = {Toh, Kang Yang and Liang, Yong Yeow and Lau, Woei Jye and Fimbres Weihs, Gustavo A.},
  month = oct,
  year = {2020},
  pages = {285},
}

@article{cespedes_improved_2025,
  title = {Improved modeling and measurements of mass transfer in spacer-filled feed channels of membrane separation processes},
  volume = {367},
  issn = {1383-5866},
  url = {https://www.sciencedirect.com/science/article/pii/S1383586625015436},
  doi = {10.1016/j.seppur.2025.132946},
  urldate = {2026-10-30},
  journal = {Separation and Purification Technology},
  author = {Cespedes, Santiago and Blankert, Bastiaan and Das, Ratul and Altmann, Thomas and Picioreanu, Cristian},
  month = sep,
  year = {2025},
  pages = {132946},
}

\end{document}